\documentclass[reprint,aps,prl,superscriptaddress,nofootinbib]{revtex4-2}
\usepackage[unicode, colorlinks=true, linkcolor=linkcolor, citecolor=linkcolor, filecolor=linkcolor,urlcolor=linkcolor, pdfusetitle]{hyperref}

\RequirePackage[top=24mm,bottom=24mm,left=22mm,right=22mm]{geometry}

\usepackage{graphicx}
\usepackage{enumitem}
\usepackage{orcidlink}
\usepackage{fancyhdr}
\usepackage{mathtools}
\usepackage{amssymb}
\usepackage{amsmath}
\usepackage{amsfonts}
\usepackage{amssymb}
\usepackage{bm}
\usepackage{soul}
\usepackage{verbatim}
\usepackage{epigraph}
\usepackage[dvipsnames]{xcolor}
\usepackage{physics}
\usepackage[normalem]{ulem}

\usepackage{tensor}
\usepackage{romannum}
\usepackage{lmodern}
\usepackage{ragged2e}
\allowdisplaybreaks
\usepackage{color}
\usepackage{framed}

\usepackage[all]{hypcap}
\usepackage[T1]{fontenc}
\usepackage[utf8]{inputenc}
\usepackage{tabularx}
\usepackage{float}
  \usepackage{cancel}
\renewcommand{\arraystretch}{1.4}

\definecolor{linkcolor}{rgb}{0.0,0.3,0.5}
\hypersetup{colorlinks, citecolor=Blue, linkcolor=blue, urlcolor=Blue}

\usepackage[table]{xcolor}
\usepackage{array}
\usepackage{colortbl}
\definecolor{tablegray}{gray}{0.75}

\def\beq{\begin{equation}}
\def\eeq{\end{equation}}
\def\ber{\begin{eqnarray}}
\def\eer{\end{eqnarray}}
\def\l{\left}
\def\r{\right}
\def\d{{\rm d}}
\def\f{\frac}
\def\mpl{m_{\rm p}}

\def \Blue {\color{Blue}}

\begin{document}

\title{Dark Matter as an Inflationary Relic in Warm Inflation}
\author{Swagat S.~Mishra\orcidlink{0000-0003-4057-145X}}
\email{swagatmishra@ibs.re.kr}
\affiliation{Cosmology,~Gravity~and~Astroparticle\,Physics Group at Center~for~Theoretical~Physics~of~the~Universe,~Institute~for~Basic~Science\,(IBS),~Daejeon,~34126,~Korea.}

\author{Umang Kumar\orcidlink{0000-0002-7220-4817}}
\email{umang.kumar\_phd21@ashoka.edu.in}
\affiliation{Department of Physics, Ashoka University,
   Rajiv Gandhi Education City, Rai, Sonipat 131\,029, Haryana, India.}

\author{Suratna Das\orcidlink{0000-0003-3785-3288}}
\email{suratna.das@ashoka.edu.in}
\affiliation{Department of Physics, Ashoka University,
   Rajiv Gandhi Education City, Rai, Sonipat 131\,029, Haryana, India.}

\author{Varun Sahni\orcidlink{0000-0002-9470-9939}}
\email{varun@iucaa.in}
\affiliation{Inter-University~Centre~for~Astronomy~and~Astrophysics, Post\,Bag\,4,~Ganeshkhind,~Pune~411\,007,~India.}

\begin{abstract}
\noindent Warm inflation is usually expected to completely deplete the inflaton condensate by dissipating its energy into radiation. We show that this expectation fails in a simple and observationally viable regime. In a strongly dissipative warm inflationary scenario, the dissipative ratio, $Q=\Upsilon/(3H)$, can fall rapidly after the end of inflation as the system approaches radiation domination, thereby suppressing further energy transfer to the thermal bath. This leads to a residual inflaton condensate, which subsequently evolves as an effectively non-dissipative scalar field. For potentials with a stable quadratic minimum, this remnant inflaton manifests as a cold dark matter component. We establish this mechanism for the minimal renormalizable potential, with a dissipative coefficient $\Upsilon\propto T^3$. In this case, current cosmological data allow strong dissipation while leaving the inflaton mass weakly constrained by inflationary observables. The observed dark matter abundance then fixes its mass to be $m \approx 0.02\,{\rm MeV}$, while larger masses overclose the Universe. The transition to matter-like scaling occurs well before BBN, avoiding a long-lived inflaton dark radiation component. Relic inflaton dark matter therefore turns the post-inflationary dynamics of warm inflation into a new late time constraint on its parameter space.
\end{abstract}

\fancypagestyle{prlfirstpage}{
  \fancyhf{}
  \fancyhead[C]{\small \texttt{\quad\quad\quad\quad\quad\quad PREPRINT LETTER~~}({\Blue June 2026})
  }
  \renewcommand{\headrulewidth}{0.6pt}
  \renewcommand{\footrulewidth}{0.0pt}
}

\maketitle

\thispagestyle{prlfirstpage}

\noindent{{\bf{\em Introduction}}}\,---\,Cosmic inflation\,\cite{Starobinsky:1980te,Guth:1980zm,Linde:1981mu,Albrecht:1982wi} provides a compelling description of the very early Universe, explaining the origin of primordial density perturbations\,\cite{Mukhanov:1981xt,Hawking:1982cz,Starobinsky:1982ee,Guth:1982ec} and setting the initial conditions for the hot Big Bang\,\cite{Linde:1984ir,Linde:1990flp, Dodelson:2003ip,Baumann:2009ds,Martin:2013tda,Baumann:2018muz,Mishra:2024axb}. In the standard picture of {\em cold inflation} (CI), however, the Universe must be reheated after inflation through the transfer of inflaton energy into relativistic degrees of freedom that eventually constitute the thermal bath of the hot Big Bang\,\cite{Albrecht:1982mp,Kolb:1990vq,Kofman:1994rk,Shtanov:1994ce,Kofman:1996mv,Kofman:1997yn,Bassett:2005xm,Amin:2014eta,Lozanov:2019jxc,Mishra:2024axb}. {\em Warm inflation} (WI) offers a different possibility: dissipative interactions continuously source a radiation bath during inflation, which can smoothly become dominant at the end of the accelerated phase\,\cite{Berera:1995ie,Berera:1995wh,Berera:1996nv,Berera:1998gx,Berera:1998px,Hall:2003zp,Graham:2009bf,Berera:2008ar,Bastero-Gil:2009sdq,Bartrum:2013fia,Rangarajan:2018tte,Kamali:2023lzq,Berera:2023liv}. In this way, WI replaces a separate reheating epoch by a dynamical transition into radiation domination\,\cite{Berera:1996fm,Berera:2023liv}.

This standard view of WI often suggests that the inflaton is efficiently depleted by its dissipative coupling to the radiation bath\,\cite{Berera:1996fm,Kamali:2023lzq,Berera:2023liv}. However, the existence of dissipation during inflation does not by itself imply that the inflaton must continue to dissipate after the accelerated phase ends. The post-inflationary efficiency of dissipation is controlled by the subsequent evolution of the dissipative term $\Upsilon(\phi,T)$ relative to the Hubble scale $H(t)$\,\cite{Rosa:2018iff}. In particular, after the end of inflation, the dimensionless ratio $Q\equiv \Upsilon/(3H)$ can fall rapidly as the system approaches radiation domination, quenching further energy transfer from the inflaton to the thermal bath. The late time fate of the inflaton condensate after WI is therefore a dynamical question, rather than an assumption. 

The key observation of this work is that a relic inflaton condensate can survive after WI when two conditions are simultaneously realized: \textit{(i)} dissipation is strong near the end of inflation, and \textit{(ii)} it becomes inefficient soon afterwards. We focus on the widely used Minimal Warm Inflation (MWI) set-up\,\cite{Berghaus:2019whh}, with dissipation $\Upsilon\propto T^3$, which arises in several low temperature dissipative realizations and provides a simple phenomenological description of strong dissipative WI. As the system approaches radiation domination, one has approximately $H\propto T^2$, and therefore $Q=\Upsilon/(3H)\propto T$. Thus, as the radiation bath cools, the dissipative channel rapidly becomes inefficient. Crucially, before this shutoff occurs, the large value of $Q$ has already substantially damped the inflaton condensate. The surviving field is therefore both weakly dissipating and sufficiently dilute. If the potential has a stable quadratic minimum, its subsequent oscillations have the time averaged equation of state $\langle w_\phi\rangle\simeq 0$, and the remnant behaves as cold dark matter\,(CDM)\,\cite{Turner:1983he,Kofman:1997yn}.

The possibility that the remnant inflaton may manifest itself as dark matter has been explored in a few contexts. In cold or hybrid inflation, incomplete inflaton decay can arise from kinematic blocking or from symmetry protected remnants\,\cite{Bastero-Gil:2015lga}. In the context of WI, the Warm Little Inflaton construction\,\cite{Bastero-Gil:2016qru} provides a symmetry based realization in which a weakly dissipative inflaton remnant first behaves as dark radiation for an extended period, including the BBN epoch, before becoming CDM prior to matter-radiation equality\,\cite{Rosa:2018iff,Levy:2020zfo}. Related scenarios have also considered dark matter production from the thermal bath during WI\,\cite{Freese:2024ogj,deSouza:2024oaz,Wang:2025duy}. The mechanism studied here is different. We do not rely on kinematic blocking or on the specific Warm Little Inflaton symmetry structure.  Instead, the inflaton condensate is strongly damped in a large $Q$ regime to begin with\,\cite{Bastero-Gil:2019gao,Das:2020xmh,Motaharfar:2021egj}, after which the rapid post-inflationary fall of $Q$ quenches further dissipation and leaves behind a surviving condensate. Its relic abundance is then determined by the same parameter space that fits inflationary observables. 

In this Letter, we demonstrate this mechanism working with the simple renormalizable potential, $V(\phi)=\f{1}{2} m^2\phi^2+\f{\lambda}{4}\phi^4$, and a MWI dissipative coefficient $\Upsilon\propto T^3$. Current cosmological data allow this model to fit the inflationary observables in the strong dissipative regime, while leaving the inflaton mass $m$ only weakly constrained. This observation plays a crucial role: the quartic coupling ($\lambda$) controls the inflationary dynamics, whereas the mass term ($m$) controls the late-time transition to the quadratic regime, and hence the relic abundance of inflaton CDM. We find that the observed CDM abundance selects $m \approx 0.02\,{\rm MeV}$, while larger masses overproduce dark matter and are thus excluded. The transition to matter-like scaling takes place well before BBN, so the remnant does not behave as a long-lived dark radiation component during nucleosynthesis, contrary to the scenario studied in \cite{Rosa:2018iff,Levy:2020zfo}. Although we focus on the renormalizable potential here, the mechanism relies only on the post-inflationary quenching of dissipation, and the presence of a stable quadratic minimum. A broader analysis of symmetry breaking potentials and other non-runaway models will be presented in a longer companion work.\\

\noindent{{\bf{\em Strong~dissipative~regime~of~WI}}}\,---\, 
We consider the standard warm inflationary set-up in a spatially flat FLRW Universe, consisting of an inflaton condensate $\phi(t)$, with potential $V(\phi)$, coupled dissipatively to an ambient radiation bath of energy density $\rho_r$ and temperature $T$. The background cosmological equations governing the expansion of the Universe, and the dynamics of the coupled inflaton and radiation system, are given by\,\cite{Kamali:2023lzq,Das:2025teu}
\ber
&&H^2 = \f{1}{3\,\mpl^2}\l(\f{1}{2} \, \dot{\phi}^2 + V(\phi) + \rho_r \r) \, ,
\label{eq:WI_H_Friedmann} \\
 &&\ddot{\phi}+3\,H\l(1+Q\r)\dot{\phi}+\f{\d V}{\d\phi} = 0 \, ,
\label{eq:WI_phi_EoM}\\
&&\dot{\rho_r}+4\,H\,\rho_r=3\,H\,Q\,\dot{\phi}^2 \, ,
\label{eq:WI_rad_EoM}
\eer
where the dimensionless ratio
\beq
Q(\phi,T) \equiv \f{\Upsilon(\phi,T)}{3\,H}
\label{eq:Q_Def}
\eeq
measures the strength of dissipative coefficient, $\Upsilon(\phi, T)$, relative to Hubble friction. The regime $Q\gg1$ corresponds to strong dissipation, in which the inflaton evolution is controlled predominantly by its coupling to the thermal bath. Note that, the density of the ambient thermal radiation  is related to its temperature \textit{via} the effective number of relativistic degrees of freedom (dof) in density, $g_*(T)$, by 
\beq
\rho_r(T) = \f{\pi^2}{30}\,g_*(T)\,T^4 \, .
\label{eq:rho_rad_T}
\eeq

We work with the minimal warm inflationary (MWI) dissipative coefficient\,\cite{Berghaus:2019whh},
\beq
\Upsilon(\phi,T) \equiv \Upsilon(T) = \f{\tilde{C}_\Upsilon}{{\cal M}^2}\,T^3 \, ,
\label{eq:Upsilon_MWI}
\eeq
where $\tilde{C}_\Upsilon$ is dimensionless and ${\cal M}$ denotes the microscopic mass scale associated with the dissipative sector. This form is the relevant phenomenological input for the mechanism studied below. During WI, the radiation bath is continuously sourced by dissipation of the inflaton, allowing the system, under suitable conditions\,\cite{Das:2020lut}, to exit inflation smoothly into radiation domination without requiring a separate reheating mechanism. 
The crucial point for our purposes is that strong dissipation during inflation does not imply continued dissipation after the end of inflation. Once the accelerated phase ends and the system approaches radiation domination, the radiation bath cools approximately as $T\propto a^{-1}$, while $H\propto T^2$. For the dissipative coefficient in Eq.\,\eqref{eq:Upsilon_MWI}, this leads to
\beq
Q \, = \, \f{\Upsilon}{3\,H} ~ \propto ~ T ~ \propto ~ \f{1}{a}\,.
\label{eq:Q_scaling_RD}
\eeq
Thus the same dissipative channel that was efficient during inflation rapidly becomes inefficient after the end of inflation. In numerical evolution this decrease begins already after $\epsilon_H=1$ and continues through the approach to radiation domination, $\epsilon_H \simeq 2$. \\

\noindent{{\bf{\em Relic~cold~dark~matter}}}\,---\,This behaviour has two important consequences. First, since $Q$ remains large near the end of inflation, the inflaton condensate continues to be strongly damped, thereby transferring most of its energy to the radiation bath. Second, once $Q$ has fallen to $Q \ll 1$, the remaining condensate no longer dissipates efficiently and evolves effectively as a non-dissipative scalar field. Hence, WI in the strong dissipative regime can leave behind a subdominant, yet surviving, inflaton remnant instead of resulting in its complete depletion into radiation. 

We define the onset of free (non-dissipative) evolution of the inflaton to be at a post-inflationary epoch when $Q_i \approx 0.01 \ll 1$. From this point onward the subsequent evolution of the remnant is governed, to an excellent approximation, by
\beq
\ddot{\phi}+3\,H\,\dot{\phi}+ \f{\d V}{\d\phi} = 0 \, ,
\label{eq:free_phi_EoM}
\eeq
with the dominant radiation component evolving independently as $\rho_r \propto a^{-4}$, up to the usual changes in the relativistic dof $g_*(T)$. If the potential has a stable quadratic minimum, then the residual oscillating condensate  eventually displays a time-averaged EoS\,\cite{Starobinsky:1982ee,Turner:1983he}, $\langle w_\phi\rangle \simeq 0$, thereby manifesting as cold dark matter in the Universe.

The above mechanism requires a potential that explains CMB observations in the strong dissipative regime of WI and leads to a graceful exit into radiation domination, while also possessing a stable quadratic minimum around which oscillations of the surviving condensate behave as CDM\,\cite{Hu:2000ke}. We demonstrate this explicitly and compute the resulting CDM abundance in the  case of a renormalizable potential as a minimal realization of this mechanism.\\

\noindent{{\bf{\em Results~for~renormalizable~potential}}}\,---\,Consider the simple renormalizable potential
\beq
V(\phi)=\f{1}{2}\,m^2\,\phi^2 + \f{\lambda}{4}\,\phi^4 \, ,
\label{eq:pot_Quad_Quart}
\eeq
together with the MWI dissipative coefficient given in Eq.\,\eqref{eq:Upsilon_MWI}. This potential has two free parameters, $\lbrace m,\,\lambda \rbrace$, with distinct physical roles. The quartic term dictates the warm inflationary dynamics in the strong dissipative regime, while the quadratic term controls the late-time oscillatory regime after inflation, and hence the relic CDM abundance. 

\begin{figure}[!t]
\centering
\includegraphics[width=0.95\linewidth]{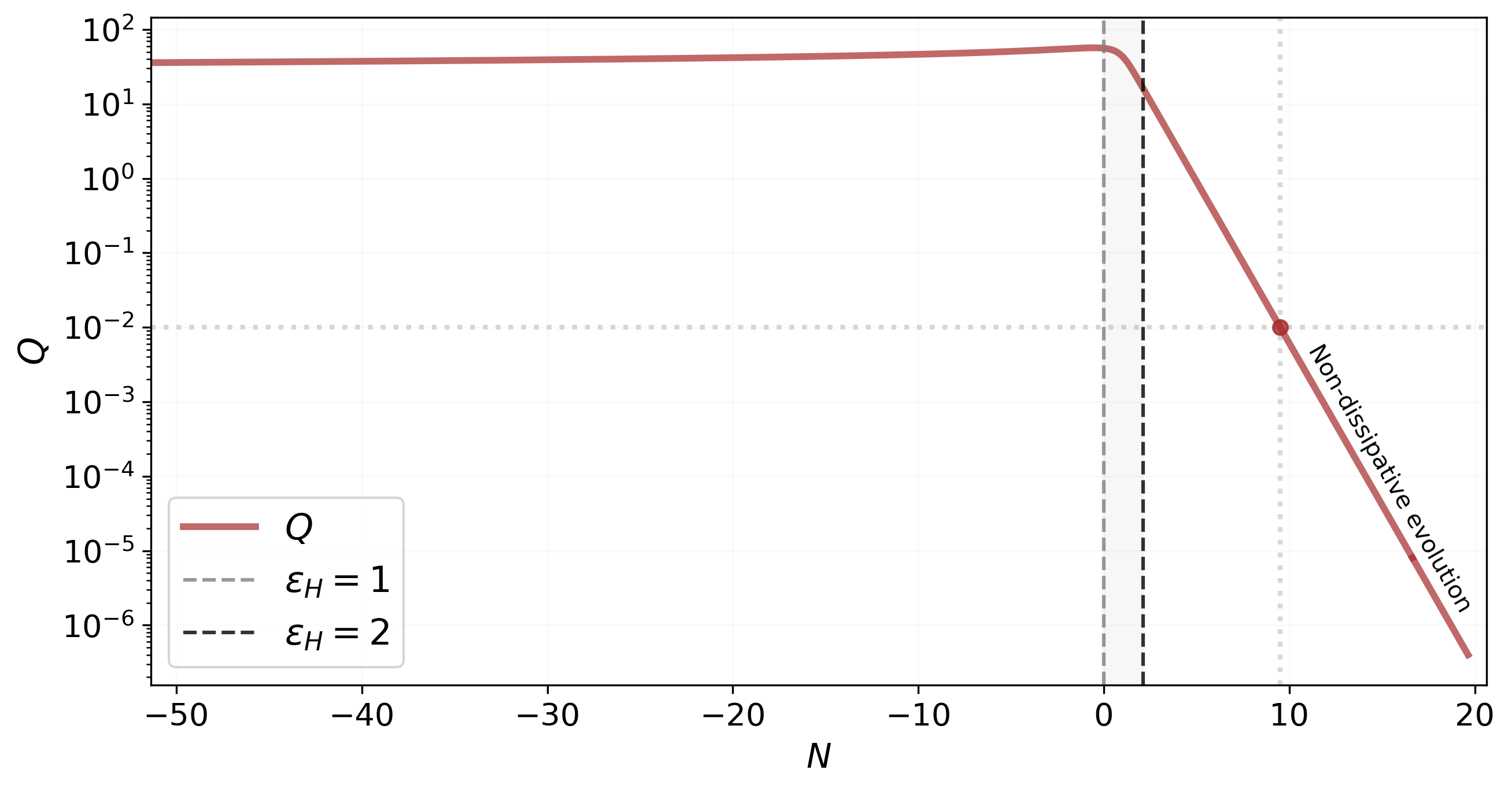}
\includegraphics[width=0.95\linewidth]{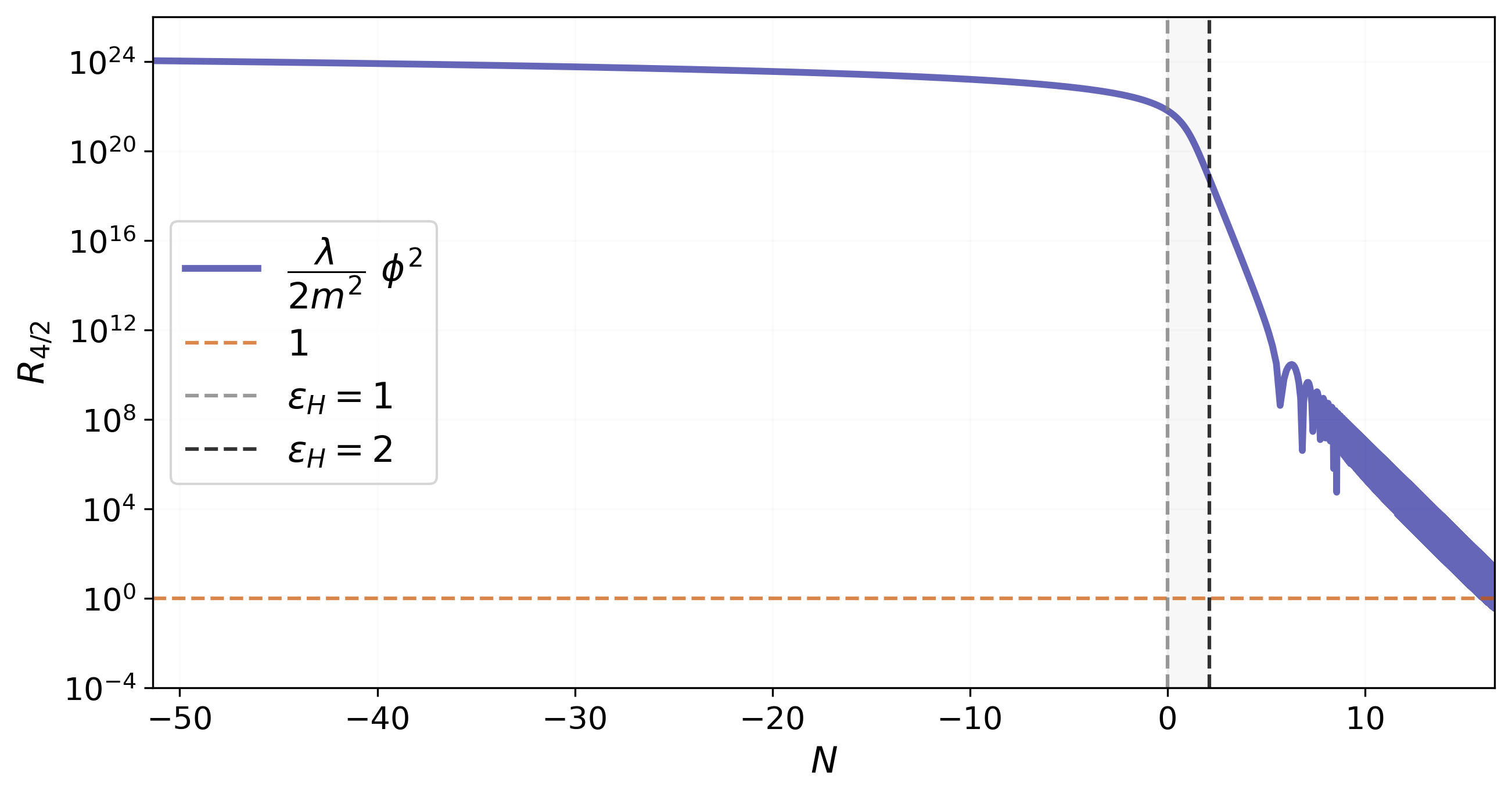}
\caption{For the renormalizable potential\,\eqref{eq:pot_Quad_Quart} with $\lambda = 1.16\times 10^{-23}$ and $m=0.017\,{\rm MeV}$. The end of inflation is denoted by $N=0$. {\bf Top panel:} Evolution of the dissipative ratio $Q$, which falls rapidly after the end of inflation. {\bf Bottom panel:} Evolution of the quartic-to-quadratic ratio $R_{4/2}\equiv \lambda\phi^2/(2m^2)$. The horizontal dashed line denotes $R_{4/2}=1$, where the quartic and quadratic contributions become equal. For $R_{4/2}>1$, the inflaton oscillations are quartic dominated and the time-averaged density scales as radiation, $\langle\rho_\phi\rangle\propto a^{-4}$; after $R_{4/2}<1$, the quadratic term dominates and the remnant scales as matter, $\langle\rho_\phi\rangle\propto a^{-3}$. The vertical dashed lines mark the end of inflation, $\epsilon_H=1$, and the onset of radiation domination, $\epsilon_H\simeq2$. }
\label{fig:rho_Q_Renorm1}
\vspace{-0.1in}
\end{figure}

We perform an MCMC analysis of this model using the Planck, ACT, SPT and DESI datasets\,\cite{Planck:2018jri,ACT:2025fju,SPT-3G:2025bzu,DESI:2025zgx}, following the standard WI prescription for the scalar power spectrum\,\cite{Ramos:2013nsa,Benetti:2016jhf,Das:2025teu} and making use of the \texttt{SWIM} numerical code\,\cite{Kumar:2026mvz}, as detailed in the {\em End Matter}. The model successfully admits an observationally viable strongly dissipative inflationary dynamics, with best-fit values, 
\beq
Q_* \simeq 35.95\, , \quad n_{_S}\simeq 0.9723\, , \quad r\simeq 7.9\times 10^{-15}\, ,
\label{eq:Renorm_Bestfit_Obs}
\eeq
and 
\beq
\lambda \simeq 1.16\times10^{-23} \, .
\label{eq:Renorm_lambda_bestfit}
\eeq
The extreme suppression of  tensor-to-scalar ratio $r$ is a characteristic consequence of WI in the strong dissipative regime\,\cite{Das:2025teu}, where the scalar spectrum receives thermal enhancement while tensor modes retain their vacuum origin\,\cite{Graham:2009bf}. Importantly, the posterior distribution, shown in the {\em End Matter}, indicates that $\lambda$ is well constrained by inflationary observables, whereas $m$ remains only weakly constrained. This is because inflation takes place in the quartic dominated regime, $\lambda\phi^4 \gg m^2\phi^2$, so the CMB fit is controlled primarily by $\lambda$. The mass parameter becomes relevant only later during the post-inflationary oscillations, when the scalar field remnant enters the quadratic regime. This is the key reason why the same model can fit the CMB while allowing for the late-time relic abundance to determine the inflaton dark matter mass. 

\begin{figure*}[!t]
\centering
\includegraphics[width=0.468\linewidth]{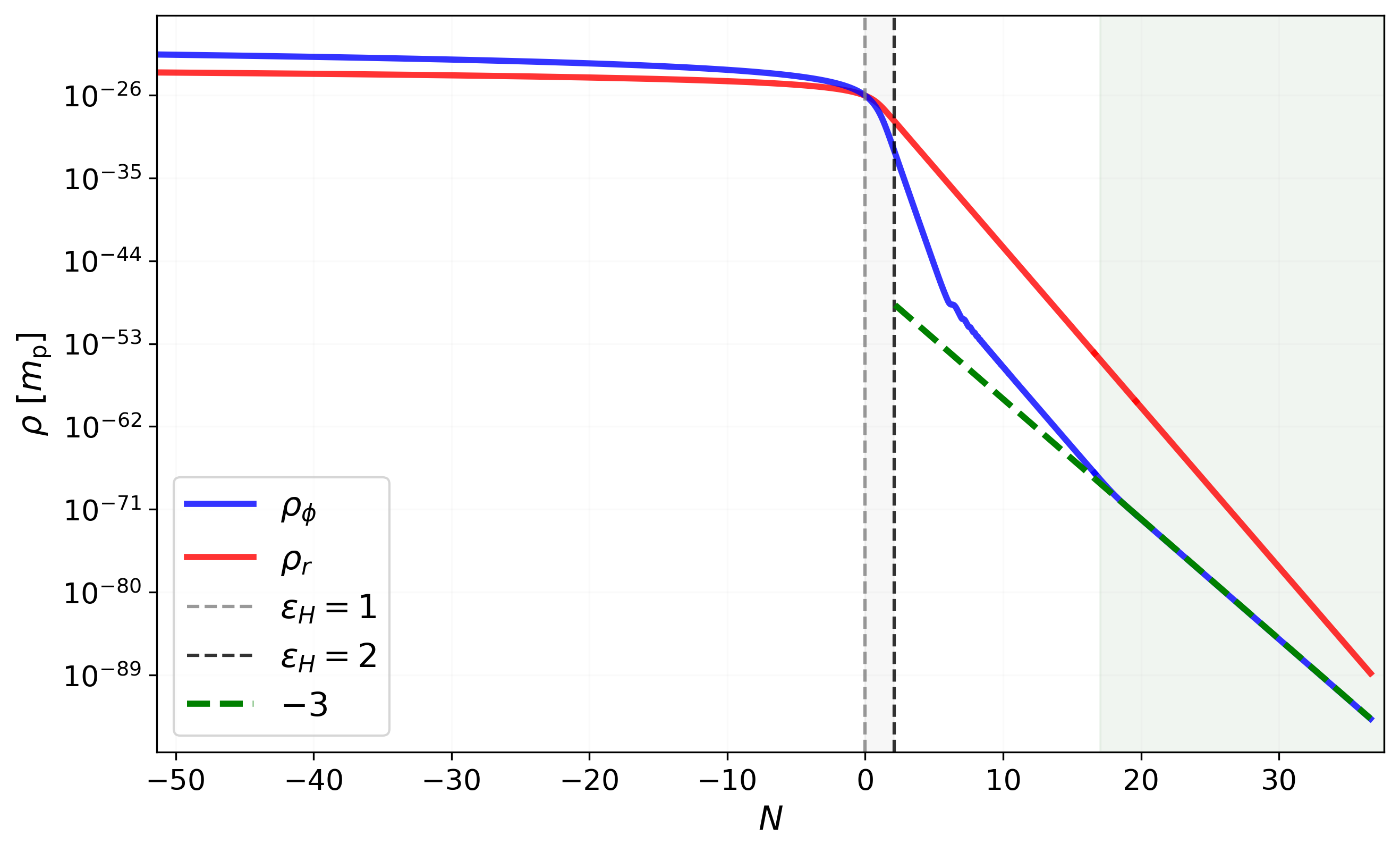}
\includegraphics[width=0.525\linewidth]{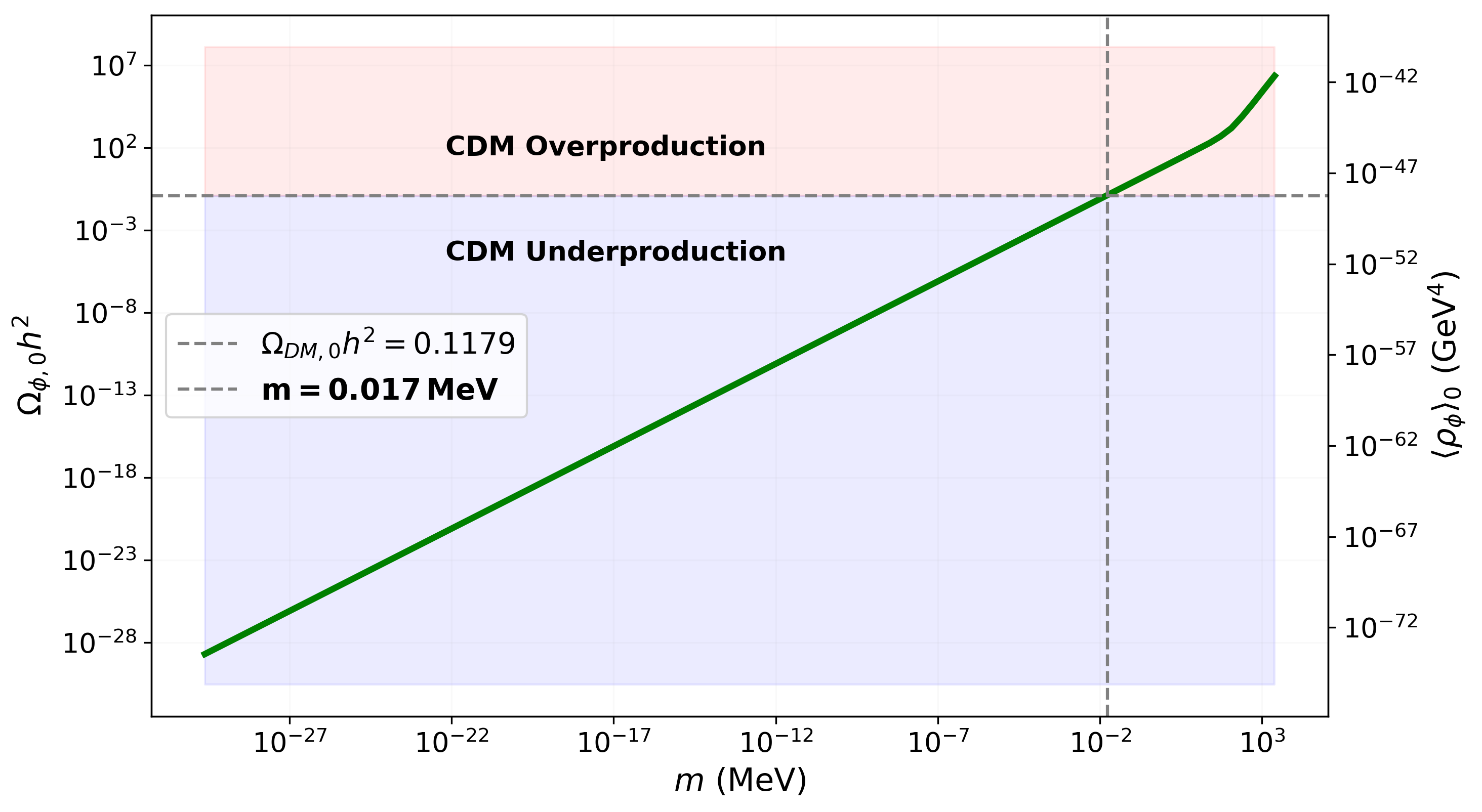}
\caption{Relic inflaton abundance for the renormalizable potential~\eqref{eq:pot_Quad_Quart} with $\lambda = 1.16\times 10^{-23}$. {\bf Left panel:} Evolution of the radiation density $\rho_r$ and the averaged inflaton density $\langle\rho_{\phi}\rangle$, plotted for $m=0.017\,{\rm MeV}$ (the end of inflation is denoted by $N=0$). After inflation, the inflaton remnant initially follows the quartic scaling $\langle\rho_\phi\rangle\propto a^{-4}$, before entering the quadratic (green shaded) regime  where $\langle\rho_\phi\rangle\propto a^{-3}$, as indicated by the dashed green line. {\bf Right panel:} Present-epoch abundance of the surviving inflaton condensate as a function of $m$, with all other model parameters fixed to their best-fit values. The horizontal dashed line denotes the observed CDM abundance. The blue and red shaded regions correspond respectively to underproduction and overproduction of inflaton cold dark matter; the red region is excluded by overclosure.}
\label{fig:Mass_DM_Renorm}
\end{figure*}

After the end of inflation, the system smoothly approaches radiation domination and the dissipative ratio $Q$ rapidly decreases, as shown in the top panel of Fig.\,\ref{fig:rho_Q_Renorm1}. As discussed before, we choose the initial epoch for computing the relic abundance to be at $Q_i=0.01$, after which the inflaton evolves effectively as a non-dissipative scalar, Eq.\,\eqref{eq:free_phi_EoM}. At this epoch, for the representative best-fit trajectory shown in the left panel of Fig.\,\ref{fig:Mass_DM_Renorm}, the inflaton condensate is already highly subdominant compared to radiation, $\rho_{ri} \gg \langle \rho_{\phi}\rangle_i$,  
as a result of the strong damping of the inflaton immediately after the end of inflation when $Q\gtrsim \mathcal{O}(10)$. This hierarchy is central, ensuring that the residual oscillating condensate, which initially behaves as dark radiation with $\langle w_\phi \rangle  \simeq 1/3$, is strongly suppressed relative to the radiation density.

The transition between the quartic and quadratic regimes is controlled by the ratio
\beq
R_{4/2} \equiv \f{\lambda \bar{\phi}^4/4}{m^2\bar{\phi}^2/2} =  \f{\lambda \, \bar{\phi}^2}{2\,m^2}\, ,
\label{eq:R_Quartic_Quadratic}
\eeq
which is shown in the bottom panel of Fig.\,\ref{fig:rho_Q_Renorm1}. Here, $\bar\phi(t)$ denotes the oscillation amplitude. Initially, the quartic term dominates with $R_{4/2}\gg1$, and the time-averaged inflaton density scales as radiation, $\langle\rho_\phi\rangle\propto a^{-4}$. When the amplitude falls below a critical value,
\beq
\bar\phi_T = \sqrt{\f{2}{\lambda}}\, m \, ,
\label{eq:phi_transition}
\eeq
one has $R_{4/2}<1$, the quadratic term dominates, and the remnant begins to scale as non-relativistic matter, $\langle\rho_\phi\rangle \propto a^{-3}$. For the parameter values yielding the observed CDM abundance, we find that this transition occurs well before the commencement of BBN, around a redshift $z_T \approx 10^{16}$ (see {\em End Matter}). Thus the oscillating remnant does not behave as a  long-lived inflaton dark radiation component during BBN.

The present-day abundance can be estimated analytically by evolving the condensate from the initial dissipation-free epoch ($Q_i = 0.01$) to the quartic-to-quadratic transition epoch $z=z_{T}$, and finally, from there to the present epoch. As derived in the {\em End Matter}, entropy conservation leads to
\beq
\langle \rho_{\phi} \rangle_0 \simeq \l[\f{g_{*s}(T_{\rm eq}) \, T_{\rm eq}^3}{g_{*s}(T_i) \,T_i^3} \r] \f{1}{(1+z_{\rm eq})^3}  \sqrt{\f{2}{\lambda}}\, \f{m}{\bar{\phi}_i} \,\langle\rho_{\phi}\rangle_i \, ,
\label{eq:rho_phi0_analytic}
\eeq
where $T_i$, $\bar\phi_i$ and $\langle\rho_{\phi}\rangle_i$ are evaluated at $Q_i=0.01$, and $g_{*s}$ denotes the  entropy degrees of freedom. This expression makes the physical dependence transparent: once the inflationary trajectory fixes $\lambda$, $T_i$, $\bar\phi_i$ and $\langle\rho_{\phi}\rangle_i$, the relic abundance is controlled primarily by the mass $m$. For a representative value of $m$ near the posterior median from our inflationary MCMC analysis, Eq.\,\eqref{eq:rho_phi0_analytic} leads to an underproduction of dark matter. However, since the quartic term predominantly governs the inflationary dynamics, the CMB fit is largely insensitive to $m$. The posterior therefore allows a broad range of $m$ without degrading the fit to inflationary observables. Using this freedom, the observed CDM abundance fixes
\beq
m\simeq 7\times 10^{-24} \, \mpl \simeq 0.017\,{\rm MeV} \, ,
\label{eq:mass_DM_Renorm}
\eeq
as shown in the right panel of  Fig.\,\ref{fig:Mass_DM_Renorm}. The above analysis has two complementary implications. First, the immediate post-inflationary phase with $Q \gtrsim \mathcal{O}(10)$ naturally suppresses the residual condensate before the dissipative channel is quenched. Therefore, the subsequent radiation-like phase of the inflaton remnant is highly subdominant, as can be seen from the left panel of Fig.\,\ref{fig:Mass_DM_Renorm}. Consequently, the transition to matter-like scaling takes place well before BBN, around $z_T \approx 10^{16}$. Second, the same post-inflationary remnant turns the otherwise weakly constrained mass parameter $m$ into a late-time observable: underproduction leaves only a subdominant relic, while overproduction excludes part of the WI parameter space, as shown in the right panel of  Fig.\,\ref{fig:Mass_DM_Renorm}. This behaviour is distinct from the Warm Little Inflaton dark matter scenario\,\cite{Rosa:2018iff}, where the remnant is protected by a specific symmetry structure and behaves as dark radiation during BBN. In our scenario, by contrast, the relic abundance follows from strong damping of the condensate followed by the rapid fall of $Q$.\\

\noindent{{\bf{\em Discussion~and~conclusions}}}\,---\,We have demonstrated that in the strong dissipative regime of WI with $\Upsilon\propto T^3$, the inflaton condensate is first efficiently damped with a large $Q$ at the end of inflation, after which the rapid post-inflationary fall of $Q$ suppresses further energy transfer to the radiation bath. This leaves a small surviving condensate which, for potentials with a stable quadratic minimum, evolves as an effectively non-dissipative scalar field and behaves as cold dark matter.

For the renormalizable potential, the quartic term controls the strongly dissipative inflationary dynamics, while the weakly constrained mass parameter controls the late-time relic abundance. Matching the observed CDM abundance selects $m \simeq 0.017\,{\rm MeV}$, whereas larger masses overproduce dark matter and are excluded. Since the transition to matter-like scaling occurs at $z_T \approx 10^{16}$, the relic inflaton remnant is already CDM well before BBN and avoids a long-lived dark radiation phase. This differs from the Warm Little Inflaton scenario, where the remnant is symmetry protected and remains radiation-like during BBN before becoming CDM\,\cite{Rosa:2018iff}.

Once the quadratic regime is reached, the inflaton remnant behaves like an ordinary oscillating scalar field CDM\,\cite{Khlopov:1985fch,Peebles:1999fz,Peebles:2000yy,Mishra:2017ehw}. For the mass scale selected here, it is far from the ultra-light fuzzy dark matter regime\,\cite{Hu:2000ke,Hui:2016ltb}, so pressure and gradient effects are negligible on structure formation scales. Since the dark matter originates from the inflaton sector, isocurvature perturbations remain an important diagnostic\,\cite{Efstathiou:1986pba}.  Although the inflationary perturbation spectrum for MWI is known and is already included in our MCMC analysis, the relevant quantity here is the perturbation of the final relic abundance, $S_{c\gamma}=\delta\ln(n_\phi/s)$, evaluated after dissipation has become inefficient. This requires propagating inflaton and radiation perturbations to the $Q_i \ll 1$ hypersurface and through the quartic-to-quadratic transition. However, the strong damping, rapid quenching of dissipation and much earlier transition to CDM make our perturbative evolution distinct from the Warm Little Inflaton case, where anticorrelated CDM isocurvature is a characteristic signature\,\cite{Rosa:2018iff}. A dedicated analysis of inflaton CDM perturbations, including the resulting isocurvature constraints, lies beyond the scope of this Letter and will be reserved for a future study.

Our analysis is deliberately restricted to the homogeneous relic abundance in the minimal renormalizable realization. A complete microphysical embedding must also address the stability of the remnant and possible residual interactions with the thermal bath after $Q\ll 1$. Nevertheless, the central conclusion is already apparent at the effective level: relic inflaton CDM provides a new late-time constraint on the same strongly dissipative WI parameter space that fits inflationary observables. Although we have focused on the renormalizable potential, our mechanism relies only on the rapid post-inflationary loss of dissipative efficiency and the presence of a stable quadratic minimum. A broader analysis, including symmetry breaking potentials\,\cite{Press:1989id}, additional non-runaway models\,\cite{Mishra:2017ehw}, perturbations and microphysical stability, will be presented in a longer companion paper.

\noindent{{\bf{\em Acknowledgments.}}}  SSM is supported by the Institute for Basic Science (IBS) as a Senior Researcher at CTPU-CGA under the project code, \texttt{IBS-R018-D3}. UK warmly acknowledges the Axis Bank PhD program at Ashoka University for the PhD fellowship provided by Axis Bank. UK also acknowledges the Ashoka University High-Performance Computing (HPC) facility for providing the computational resources used in this work. VS thanks the Anusandhan National Research Foundation (ANRF), India, for the National Science Chair Professorship which provided partial funding for this work. For the purpose of open access, the authors have applied a CC BY public copyright license to any Author Accepted Manuscript version arising.

\bibliographystyle{apsrev4-2}

\bibliography{Bibliography_Letter}

\onecolumngrid

\newpage 

\appendix

\begin{center}
{\large \bf End Matter}
\end{center}
\subsubsection*{Inflationary observables and numerical set-up}
In the slow-roll regime of WI, the Hubble slow-roll parameter is approximately suppressed by the dissipative friction as
\beq
\epsilon_H \simeq \f{\epsilon_{_V}}{1+Q}\,; \qquad \text{where} \qquad \epsilon_V=\f{\mpl^2}{2}\l(\f{1}{V}\,\f{\d V}{\d\phi}\r)^2 \, .
\label{eq:epsH_WI_app}
\eeq
Thus, even relatively steep potentials can support WI\,\cite{Benetti:2016jhf,Das:2020xmh,Das:2025teu} in the strong dissipative regime with $Q\gg1$. In our numerical analysis, however, we solve the full background equations, Eqs.\,\eqref{eq:WI_H_Friedmann}\,--\,\eqref{eq:WI_rad_EoM}, rather than relying only on the slow-roll approximation. Inflation ends when $\epsilon_H=1$, while radiation domination is reached shortly afterwards when $\epsilon_H\simeq2$. (In the numerical implementation,  we identify the onset of radiation domination with $\epsilon_H \simeq 1.99$, which is indistinguishable from $\epsilon_H = 2$ at the level relevant for the present analysis.)

Since the inflaton is coupled to the radiation bath, the scalar power spectrum in WI differs from its CI form. Under the standard slow-roll approximation, it can be written as\,\cite{Ramos:2013nsa,Benetti:2016jhf}
\beq
{\cal P}_{\mathcal R}= {\cal P}_{\mathcal R}^{\rm CI} \times {\mathcal F}(Q)\, ; \qquad {\cal P}_{\mathcal R}^{\rm CI} =\l(\f{H^2}{2\pi\,\dot{\phi}}\r)^2 \, ,
\label{eq:PS_WI_app}
\eeq
where
\beq
{\mathcal F}(Q)= \l[ 1 + 2\, n_{\rm BE}(T) + \l(\f{2\sqrt{3}\pi \,Q}{\sqrt{3 + 4\pi \,Q}}\r)\l(\f{T}{H}\r)\r]G(Q)\, .
\label{eq:FQ_app}
\eeq
Here $n_{\rm BE}(T)$ is the Bose-Einstein occupation number of inflaton fluctuations when these are thermalized with the radiation bath, while $G(Q)$ accounts for the coupled evolution of inflaton and radiation perturbations.
An approximated analytic form of $G(Q)$ for MWI with runaway exponential potential was proposed in\,\cite{Das:2020xmh} which can be written as 
\beq
G(Q)=\frac{1+6.12 \, Q^{2.73}}{(1 + 6.96 \, Q^{0.78})^{0.72}}+\frac{0.01 \, Q^{4.61}(1+4.82\times 10^{-6} \, Q^{3.12})}{(1+6.83\times 10^{-13} \, Q^{4.12})^2}\,.
\label{eq:GQ_analytic_app}
\eeq
As the $G(Q)$ depends very weakly on the form of the potential, one can use this form of $G(Q)$ for the renormalizable potential studied in this work. However, we make use of \texttt{SWIM}\,\cite{Kumar:2024hju,Kumar:2026mvz} to generate the numerical form of the $G(Q)$ function (see Refs.\,\cite{Montefalcone:2023pvh,Rodrigues:2025neh} for other numerical codes in the literature). 

The amplitude of the scalar power spectrum is fixed at a specific pivot scale, $k_*$, at a time when this scale crosses the horizon during inflation. 
The scalar spectral index $n_{_S}$ and its running $\alpha_{_S}$ can be determined at the horizon crossing of the pivot scale $k_*$ as
\beq
n_{_S}-1 = \f{\d\ln {\mathcal P}_{\mathcal R}(k/k_*)}{\d\ln(k/k_*)}\bigg\vert_{k = k_*}\,; \qquad 
\alpha_{_S} =\f{\d n_{_S}(k/k_*)}{\d\ln(k/k_*)}\bigg\vert_{k =k_*} \, .
\label{eq:nS_alphaS_app}
\eeq
On the other hand, as the tensor perturbations in WI do not couple to the radiation bath, the primordial tensor spectrum takes the same form in WI as in CI,
\beq
{\mathcal P}_T \equiv {\mathcal P}_T^{\rm CI}=\f{2}{\pi^2}\l(\f{H}{\mpl}\r)^2\,.
\label{eq:PT_WI_app}
\eeq
The tensor-to-scalar ratio thus becomes,
\beq
r\equiv \f{{\mathcal P}_T(k/k_*)}{{\mathcal P}_{\mathcal R}(k/k_*)}\bigg\vert_{k = k_*} = \f{r^{\rm CI}}{{\mathcal F}(Q)}\,,
\label{eq:r_WI_app}
\eeq
which is strongly suppressed with respect to its CI counterpart. 

\begin{table}[!t]
\begin{minipage}{0.485\textwidth}
\arrayrulecolor{tablegray}
\setlength{\arrayrulewidth}{1pt}
\renewcommand{\arraystretch}{2}
\setlength{\tabcolsep}{8pt}
\begin{tabular}{|c|c|}
\hline
 {\bf Observables} &    {\bf Best-fit Values} 
\tabularnewline
\hline
   $\ln(10^{10}A_s)$ &    $3.0667$ 
\tabularnewline
\hline
  $n_{_S}$ &    $0.9723$ 
\tabularnewline
\hline
   $r$ &    $7.9 \times 10^{-15}$ 
\tabularnewline
\hline
   $z_{\rm eq}$ &    $3352.244$     
\tabularnewline
\hline
   $H_0$ &    $68.23~{\rm km}\,{\rm s}^{-1}\,{\rm Mpc}^{-1}$    
\tabularnewline
\hline
   $\Omega_m\,h^2$ &    $0.14$  
\tabularnewline
\hline
\end{tabular}
\end{minipage}
\hfill
\begin{minipage}{0.485\textwidth}
\arrayrulecolor{tablegray}
\setlength{\arrayrulewidth}{1pt}
\renewcommand{\arraystretch}{2}
\setlength{\tabcolsep}{10pt}
\begin{tabular}{|c|c|c|}
\hline
 {\bf Parameters} &  {\bf Best-fit Values} 
\tabularnewline
\hline
  $\bm{\lambda}$ &   $\bm{1.1565 \times 10^{-23}}$  
 \tabularnewline
\hline
  $\bm{m}$ &   $\bm{-}$ 
 \tabularnewline
\hline
  $g_*$ &   $227.752$ 
 \tabularnewline
\hline
  $C_\Upsilon \equiv \tilde{C}_\Upsilon/{\cal M}^2$  &   $1.1584 \times 10^{10}\,m_{\rm p}^{-2}$ 
\tabularnewline
\hline
  $\phi_{*} $  &   $3.0448\,m_{\rm p}$ 
\tabularnewline
\hline
  $\bm{Q_{*}}$  &   $\bm{35.9513}$
\tabularnewline
\hline
\end{tabular}
\end{minipage}
\caption{Best-fit inflationary  parameters and cosmological  observables at the pivot scale for the renormalizable WI model with the MWI dissipative coefficient  given in Eq.\,\eqref{eq:Upsilon_MWI}. The {\em dash} for $m$ indicates that the mass is weakly constrained by inflationary observables and is fixed instead by the late-time relic abundance.}
\label{tab:bestfit_observables}
\end{table}

\begin{figure*}[t] 
\centering
\begin{minipage}{0.6\textwidth}
    \centering
\includegraphics[width=0.9\linewidth]{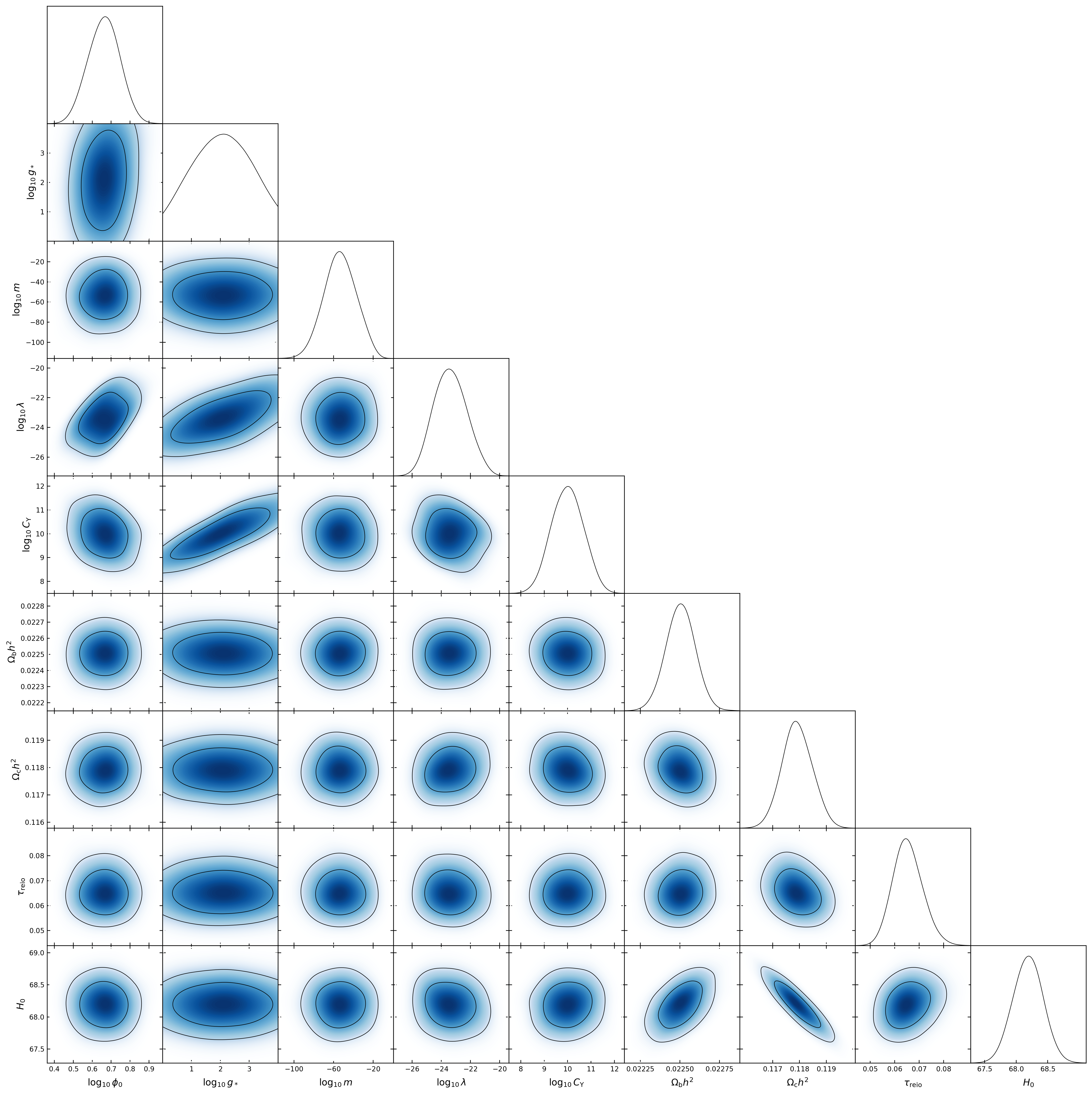}
\caption{Posterior distribution of inflationary (and other cosmological)  parameters for the renormalizable WI model with potential~\eqref{eq:pot_Quad_Quart} and dissipative coefficient~\eqref{eq:Upsilon_MWI}. Here, $C_\Upsilon = \tilde{C}_\Upsilon/{\cal M}^2$. The quartic coupling $\lambda$ is constrained by the data, while the mass parameter $m$ (expressed in the unit of $\mpl$) remains weakly constrained.}
\label{fig:contour_plot_RP_T3_1}
\end{minipage}
\begin{minipage}{0.38\textwidth}
\vspace{-0.2in}
    \centering
    \makeatletter
    \def\@captype{table}
    \makeatother
\arrayrulecolor{tablegray}
\setlength{\arrayrulewidth}{1pt}
\renewcommand{\arraystretch}{2.2}
\setlength{\tabcolsep}{10pt}
\vspace{-0.2in}
\begin{tabular}{|c|c|}
\hline
  {\bf Variables} &   {\bf Values at $\bm{Q = 0.01}$} 
\tabularnewline
\hline
  $\bm{\bar{\phi}_i}$ &   $\bm{1.7454 \times 10^{-8}\,m_{\rm p}}$ 
\tabularnewline
\hline 
  $\bm{T_i}$ &   $\bm{1.2943 \times 10^{-11}\,m_{\rm p}}$  
\tabularnewline
\hline 
  $H_i$  &   $8.3715\times 10^{-22} \, m_{\rm p}$ 
\tabularnewline
\hline 
  $\Upsilon_i$  &    $2.5115\times 10^{-23}\, m_{\rm p}$ 
\tabularnewline
\hline 
  $\bm{\langle \rho_{\phi}\rangle_i}$ &   $\bm{2.576 \times 10^{-55}\,m^4_{\rm p}}$ 
\tabularnewline
\hline 
  $\bm{\rho_{ri}} \gg \bm{\langle \rho_{\phi}\rangle_i}$ &   $2.1025 \times 10^{-42}\,m^4_{\rm p}$ 
\tabularnewline
\hline
\end{tabular}
\caption{Quantities at the post-inflationary epoch when $Q_i=0.01$, which we use as the onset of effectively non-dissipative evolution of the inflaton.}
\label{tab:initial_Q0p01}
\end{minipage}
\end{figure*}

\subsubsection*{Analytical estimate of the relic CDM abundance}
We now provide an analytical formula used in computing the present-epoch abundance of the remnant inflaton CDM condensate. The matching is performed at a post-inflationary epoch $t =t_i$, defined by $Q_i=0.01$. This choice is an operational definition of the onset of effectively non-dissipative evolution: at this point the dissipative friction is already only at the percent level compared to Hubble friction. The oscillation amplitude $\bar{\phi}_i$ and time-averaged density $\langle\rho_{\phi}\rangle_i$ are extracted from the numerical solution at this epoch.

For the renormalizable potential in Eq.\,\eqref{eq:pot_Quad_Quart}, the relative importance of the quartic and quadratic contributions is measured by
\beq
R_{4/2} \equiv \f{\lambda\bar{\phi}^4/4}{m^2\bar{\phi}^2/2} = \f{\lambda\bar{\phi}^2}{2\,m^2}\, .
\label{eq:R42_app}
\eeq
At $Q_i=0.01$, the remnant is in the quartic dominated regime, $R_{4/2} \gg 1$, and therefore
\beq
\langle \rho_\phi\rangle \propto a^{-4} \qquad \Longrightarrow \qquad \bar{\phi}\propto \f{1}{a}\, .
\label{eq:quartic_scaling_app}
\eeq
The transition to matter-like scaling occurs when $R_{4/2}=1$, namely when the oscillation amplitude reaches\
\beq
\bar\phi_T=\sqrt{\f{2}{\lambda}}\,m \, .
\label{eq:phi_transition_app}
\eeq
Using $\bar\phi\propto a^{-1}$ in the quartic regime, the scale factor at the transition satisfies
\beq
\f{a_i}{a_T} ~\equiv ~\f{1+z_T}{1+z_i} ~= ~\f{\bar{\phi}_T}{\bar{\phi}_i} = \sqrt{\f{2}{\lambda}}\,\f{m}{\bar{\phi}_i} \, .
\label{eq:ai_aT_app}
\eeq
Entropy conservation  between $t_i$ and matter-radiation equality leads to
\beq
1+z_i = (1+z_{\rm eq}) \l[\f{g_{*s}(T_i)}{g_{*s}(T_{\rm eq})}\r]^{1/3} \f{T_i}{T_{\rm eq}}\, .
\label{eq:zi_entropy_app}
\eeq
Using Eq.\,\eqref{eq:zi_entropy_app} in Eq.\,\eqref{eq:ai_aT_app}, we obtain
\beq
1+z_T = (1+z_{\rm eq}) \l[\f{g_{*s}(T_i)}{g_{*s}(T_{\rm eq})}\r]^{1/3} \f{T_i}{T_{\rm eq}} \sqrt{\f{2}{\lambda}}\, \f{m}{\bar{\phi}_i}\, .
\label{eq:zT_app}
\eeq
For the parameter values used in Fig.~\ref{fig:Mass_DM_Renorm}, this yields $z_T \sim 10^{16}$. The time-averaged inflaton density at the transition is therefore
\beq
\langle\rho_{\phi}\rangle_T = \langle\rho_{\phi}\rangle_i \l(\f{a_i}{a_T}\r)^4\, .
\label{eq:rho_phi_T_app}
\eeq
After this point the quadratic term dominates and the condensate scales as non-relativistic matter, $\langle\rho_\phi\rangle\propto a^{-3}$,
hence
\beq
\langle\rho_{\phi}\rangle_0 = \langle\rho_{\phi}\rangle_T \l(\f{a_T}{a_0}\r)^3 =  \langle\rho_{\phi}\rangle_i \l(\f{1+z_T}{1+z_i}\r) \l(\f{1}{1+z_i}\r)^3 \, .
\label{eq:rho_phi0_steps_app}
\eeq
 Using Eqs.\,\eqref{eq:ai_aT_app}~and~\eqref{eq:zi_entropy_app} in Eq.\,\eqref{eq:rho_phi0_steps_app}, we obtain
\beq
\langle\rho_{\phi}\rangle_0 =
\l[\f{g_{*s}(T_{\rm eq})\,T_{\rm eq}^3} {g_{*s}(T_i)\,T_i^3} \r] \f{1}{(1+z_{\rm eq})^3} \sqrt{\f{2}{\lambda}}\, \f{m}{\bar{\phi}_i}\, \langle\rho_{\phi}\rangle_i\, ,
\label{eq:rho_phi0_master_app}
\eeq
which is the master formula used in the main text. It shows explicitly that, once the inflationary trajectory fixes $\lambda$, $T_i$, $\bar{\phi}_i$ and $\langle\rho_{\phi}\rangle_i$, the present-epoch abundance is controlled primarily by the mass parameter $m$. The corresponding density parameter is given by
\beq
\langle \Omega_{\phi}\rangle_0= \f{\langle\rho_{\phi}\rangle_0}{\rho_{c0}}\, ,
\label{eq:Omega_phi_h2_app}
\eeq
where $\rho_{c0}$ is the present critical density. For a near-median value of the mass obtained from the inflationary MCMC posterior, $m\simeq 5 \times10^{-51}\,\mpl$, Eq.\,\eqref{eq:rho_phi0_master_app} leads to a severe underproduction of dark matter, namely,
\beq
\langle\rho_{\phi}\rangle_0 \approx 6 \times 10^{-61}\,{\rm GeV}^4 \simeq \mathcal{O}(10^{-13})\,\rho_{{\rm DM}0}\, .
\label{eq:underproduction_app}
\eeq
However, the posterior in Fig.\,\ref{fig:contour_plot_RP_T3_1} allows a broad range of $m$ without compromising the inflationary fit. Requiring Eq.\,\eqref{eq:rho_phi0_master_app} to reproduce the observed CDM abundance yields
\beq
m \simeq 7 \times 10^{-24}\,\mpl \simeq 0.017\,{\rm MeV}\, .
\label{eq:m_DM_app}
\eeq
Masses below this value underproduce the observed CDM abundance, while larger masses overproduce it and are excluded by overclosure. This is the origin of the shaded blue and red regions shown in the right panel of Fig.\,\ref{fig:Mass_DM_Renorm}.



\end{document}